\documentclass{revtex4}
\usepackage[ansinew]{inputenc}
\usepackage{color,graphicx,calc,url,amsfonts,amsmath,amssymb}

\newcommand{\linbo}{$\rm LiNbO_{3}~$}

\begin{document}
\title{Out-of-phase mixed holographic gratings : a quantative analysis}
\author{Martin Fally}
\affiliation{Faculty of Physics, Nonlinear Physics, University of Vienna, Boltzmanngasse 5, A-1090 Wien, Austria}
\email{martin.fally@univie.ac.at}
\homepage{http://nlp.univie.ac.at}
\author{Mostafa A. Ellabban}
\affiliation{Faculty of Physics, Nonlinear Physics, University of Vienna, Boltzmanngasse 5, A-1090 Wien, Austria}
\affiliation{Physics Department, Faculty of Science, Tanta University, Tanta 31527, Egypt}
\author{Irena Dreven\v{s}ek-Olenik}
\affiliation{Faculty of Mathematics and Physics, University of Ljubljana, Jadranska 19, SI-1001 Ljubljana, Slovenia}
\affiliation{J. Stefan Institute, Jamova 39, SI-1001 Ljubljana,  Slovenia}
\date{\today, \bf CORRECTION}
\begin{abstract}
We show, that by performing a simultaneous analysis of the angular dependencies of the $\pm$ first and the zeroth diffraction orders of mixed holographic gratings, each of the relevant parameters can be obtained: the strength of  the phase  grating and the amplitude  grating, respectively, as well as a potential phase between them. Experiments on a pure lithium niobate crystal are used to demonstrate the applicability of  the analysis.
\end{abstract}
\maketitle
    \section{Introduction}
Recently, volume holographic gratings with a modulation of both, the absorption coefficient and the refractive index,  have attracted attention in various materials such as silver halide emulsions \cite{Carretero-ol01,Neipp-jpd02,Neipp-oex02}, doped garnet crystals \cite{Ellabban-oex06} or materials with colloidal color centers\cite{Shcheulin-os07} . The simplest theoretical description (two-wave-coupling theory)  of light propagation in an isotropic medium with a periodic modulation of the (complex) dielectric constant was already given long ago by Kogelnik \cite{Kogelnik-bell69}. Considering periodic phase and amplitude modulations, the grating types are treated to be \emph{in phase}. Later Guibelalde generalized the equations to be valid for \emph{out-of-phase}  gratings \cite{Guibelalde-oqe84}. The quantity of major interest usually is the (first order) diffraction efficiency $\eta_{1}$, defined as the ratio of powers between the  diffracted beam and the incoming beam. For the case of  high diffraction efficiencies (above 50\%) or even for overmodulated gratings \cite{Neipp-joa01,Neipp-oex02,Gallego-oc03}, the so called 'transmission efficiency' $\eta_0$, i.e., more correctly termed as  zero order diffraction efficiency, was also employed  for characterization of the grating parameters. It was suggested, that by measuring the diffraction and transmission efficiency it is possible to evaluate the refractive-index modulation $n_1$ and the absorption constant modulation $\alpha_1$ if one assumes in-phase gratings \cite{Carretero-ol01}.

In this article we show how the shape of the angular dependencies for the $\pm$ first and zero order diffraction efficiencies depend characteristically on the parameters $n_1, \alpha_1$ and the phase $\varphi$ between them. We generalize the formulae given in Ref. \cite{Carretero-ol01} to the case of out-of-phase gratings and demonstrate at two experimental examples that the analysis is applicable. This is important, as up to now the evaluation of mixed gratings including a phase was only conducted by beam-coupling experiments, an interferometric technique which is more demanding from an experimental point of view.
    \section{Diffraction efficiencies of zero and $\pm$ first order}
\begin{figure}[h]
  \includegraphics[width=\columnwidth/2]{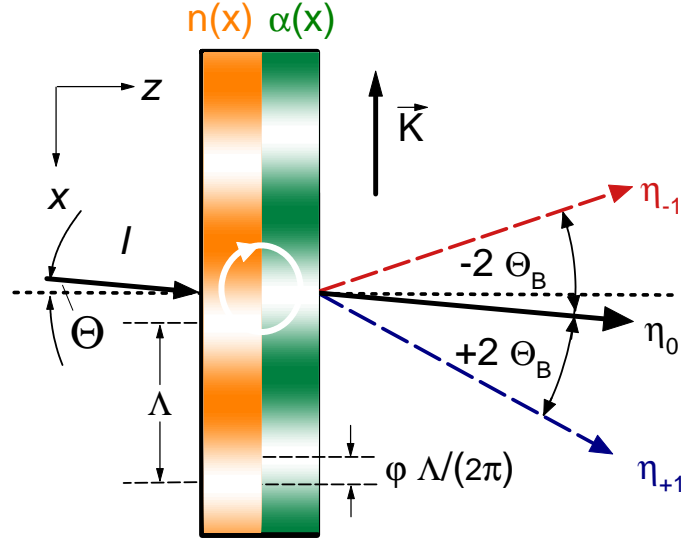}
\caption{\label{fig:setup} Schematic of the setup. The angular dependence of the zeroth $\eta_0(\Theta)$ and $\pm$ first diffraction orders $\eta_{\pm1}(\Theta)$ from a mixed grating is measured by rotating the sample around an axis perpendicular to the grating vector $\vec{K}$ . $\Theta$ denotes angles outside the medium. Note, that we are within the thick grating regime, where only {\em two} beams are propagating simultaneously: the zero order together with {\em either} the $-1^{st}$ {\em or} the $+1^{st}$ order.}
\end{figure}

According to Refs. \cite{Kogelnik-bell69,Guibelalde-oqe84} a plane wave propagating in a (thick) medium with a one dimensional periodic complex dielectric constant, composed of its real part $n(x)=n_0+n_1 \cos{(Kx)}$ and imaginary part $\alpha(x)=\alpha_0+\alpha_1 \cos{(Kx+\varphi)}$,  yields outgoing complex electric field amplitudes for the (zero order) forward diffracted  $\hat{R}_0$ and (first order) diffracted $\hat{R}_{\pm1}$  waves. These depend characteristically on the following parameters: the mean absorption constant $\alpha_0$, the thickness $d$ of the grating, the dephasing $\vartheta$ due to the deviation from Bragg's law  and the complex coupling constant $\kappa^{\pm}=n_1\pi/\lambda-i\alpha_1/2 e^{\pm i\varphi}=\kappa_1-i\kappa_2e^{\pm i\varphi}$ .  Further, $K$ denotes the spatial frequency of the grating, $n_0$  the mean refractive index of the medium, and $\varphi$ a possible phase shift between the refractive-index and absorption grating.
The goal of an experiment is to extract the grating parameters $n_1,\alpha_1, \varphi$ by varying the dephasing, e.g., through measuring the angular response of $\eta_0=\hat{R}_0\hat{R}_0^*/I$ and $\eta_{\pm 1}=\hat{R}_{\pm1}\hat{R}_{\pm1}^*/I$ where $^*$ denotes the complex conjugate and $I$ the incident intensity.
For simplicity in calculations and as the most often used experimental setup we assume a symmetrical geometry, i.e., that the grating vector and the surface normal are mutually perpendicular. A schematic of the setup is shown in Figure \ref{fig:setup}.

 Slightly adapting the convenient notation from Ref. \cite{Carretero-ol01} the efficiencies for transmission gratings can easily be calculated to yield
\begin{eqnarray}\label{eq:DE}
\eta_{\pm1}(\theta)&=&2 A(\theta)
\frac{\kappa_1^2+\kappa_2^2\pm2\kappa_1\kappa_2\sin{\varphi}}{z}
\left(\cosh{\left[z^{1/2}D\cos{\psi}\right]}-\cos{\left[z^{1/2}D\sin{\psi}\right]}
\right)\\
\eta_0(\theta)&=&\frac{A(\theta)}{2z}
\left(
(z+\vartheta^2) \cosh{\left[z^{1/2}D\cos{\psi}\right]}+(z-\vartheta^2) \cos{\left[z^{1/2}D\sin{\psi}\right]}\right. \nonumber \\
 & +& \left.2\frac{\cos{\varphi}}{|\cos{\varphi}|}\vartheta z^{1/2} \left\{\sin{\psi}\sinh{\left[z^{1/2}D\cos{\psi}\right]}-\cos{\psi}\sin{\left[z^{1/2}D\sin{\psi}\right]}
\right\}\label{eq:TE}
\right)
\end{eqnarray}
with the abbreviations $A(\theta)=\exp{\{-2\alpha_0 D\}}$,  $D=d/\cos{\theta}$ and
\begin{eqnarray}
\vartheta&=&K (\sin{\theta}-\sin{\theta_B})\\
z&=&\left\{[\vartheta^2+4 (\kappa_1^2-\kappa_2^2)]^2+[8\kappa_1\kappa_2\cos{\varphi}]^2\right\}^{1/2}\\
2\psi&=& \arccos{\left(-\frac{\vartheta^2+4(\kappa_1^2-\kappa_2^2)}{z}\right)}.
\end{eqnarray}
Here,   $\theta_B$ denotes the Bragg angle (inside the medium).  Equations (\ref{eq:DE}) and (\ref{eq:TE}) are valid for $\theta\geq0$; for $\theta\leq0$ the angles and phase-shifts are replaced by their negative values, i.e.,  $\eta_{\pm1}(-\theta)=\eta_{\mp1}(\theta)$ and  $\eta_{\pm1}(-\varphi)=\eta_{\mp1}(\varphi)$.
Note, that Equation  (\ref{eq:DE}) is identical to Equation (11) from Ref. \cite{Guibelalde-oqe84}.
Employing Equations  (\ref{eq:DE}) and  (\ref{eq:TE}) we now study the particular case of $\alpha_0 d=1$ and  $\kappa_2=\alpha_0/2$, i.e., maximal grating strength for the amplitude contribution \cite{Kogelnik-bell69}. We vary the strength of the phase grating between $\kappa_1=\kappa_2/4,\kappa_2, 4 \kappa_2$ with different phase angles $\varphi=0,\pi/4,\pi/2,3\pi/4$ between the grating types. The angular dependencies of the zero order and $\pm$ first order diffraction efficiencies are depicted in  Figure 1(a)-(d).
\begin{figure}
\includegraphics[width=\columnwidth]{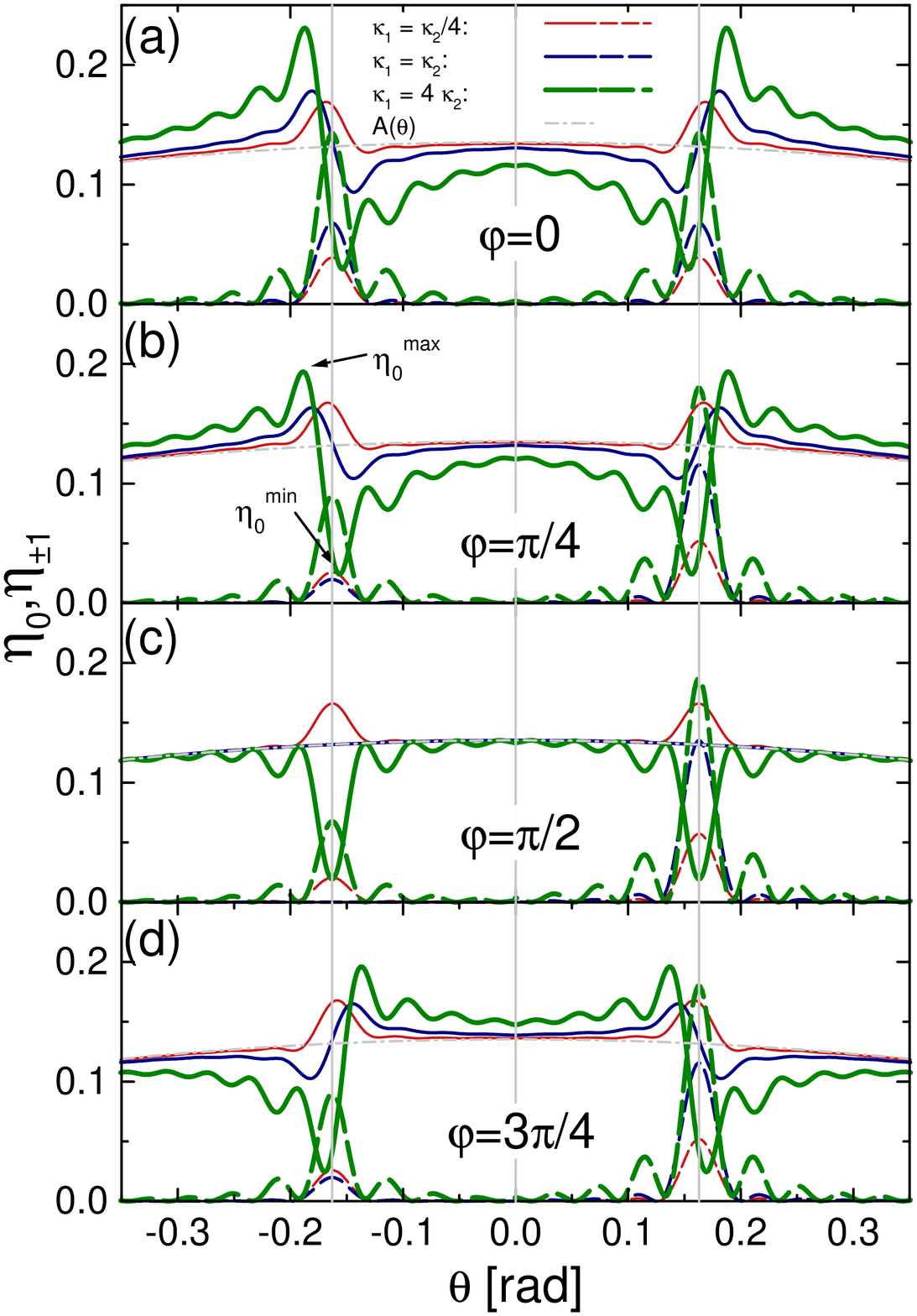}
\caption{\label{fig:difftrans4}Angular dependence of the zeroth and first orders diffraction efficiencies for increasing grating strength of the phase grating contribution. For all graphs:  $\alpha_0=2.5\times 10^4$,$\alpha_0 d=1$, $\kappa_2=\alpha_0/2$. Red thin lines: $\kappa_1=\kappa_2/4$, blue lines: $\kappa_1=\kappa_2$ and green thick lines:$\kappa_1=4 \kappa_2$ with $\varphi=0$  (a), $\varphi=\pi/4$ (b), $\varphi=\pi/2$ (c),  and $\varphi=3\pi/4$. The dash-dot line indicates the mean absorption curve $A(\theta)$ . $\theta=0,\pm\theta_B$ are marked by vertical lines. Note, that for $\varphi=\pi/2,\kappa_1=\kappa2$ the minus first order Bragg peak completely disappears and the zero order peaks, too  (shown in (c), blue lines)}
\end{figure}

At this point let us summarize the main characteristic features occurring in the diffraction efficiencies at the example for $\varphi=\pi/4$ to obtain a qualitative understanding of the curve shapes and their dependency on the ratio of $\kappa_1/\kappa_2$:
\begin{itemize}
\item Zero order diffraction efficiency $\eta_0(\theta)$
\begin{itemize}
\item The curves are symmetric with respect to normal incidence, i.e., $\theta=0$.
\item Neither the minima nor the maxima of the curve are located at the Bragg angle, except for $\kappa_1=0$ or $\kappa_2=0$ or $\varphi=\pi/2$. In general the position and the height of the minima or maxima depend in a complex way on $\kappa_1,\kappa_2,\varphi$ and even the mean absorption constant $\alpha_0$ (see discussion for $\alpha_0d\gg1$).
\item For $\kappa_1<\kappa_2$ the curve at  the Bragg angle  extends more to the region above the mean absorption curve (dash-dot line, first term in Equations (\ref{eq:DE}) and (\ref{eq:TE}) ) than below, i.e.,  as a simple approximation $\eta_0^{max}+\eta_0^{min}>2A(\theta_B)$. The same is true vice versa for $\kappa_1>\kappa_2$
\item Note, that for $|\theta|\ll\theta_B$ the curve resides below the mean absorption curve, for $|\theta|\gg\theta_B$ above
\end{itemize}
\item Diffraction efficiency
\begin{itemize}
\item The maximum value of the diffraction efficiency differs for $\eta_{-1}(\theta_B)$ and $\eta_{+1}(\theta_B)$; in our case  $\eta_{-1}(\theta_B)< \eta_{+1}(\theta_B)$.
\item The curves are symmetric with respect to $\theta_B$, i.e., $\eta_{1}(\theta_B+x)=\eta_{1}(\theta_B-x)$  except for their different mean absorption $A(\theta_B\pm x)$.
\item Note, that despite $\kappa_1<\kappa_2$ the diffraction efficiency $\eta_{-1}(\theta_B,\kappa_2/4)>\eta_{-1}(\theta_B,\kappa_2)$  for the minus first diffraction order, whereas it is vice versa for the plus first diffraction order, i.e., $\eta_{+1}(\theta_B,\kappa_2/4)<\eta_{+1}(\theta_B,\kappa_2)$
\end{itemize}
\end{itemize}
Next we would like to point out the difference between the curves for various $\varphi$ values. Figure \ref{fig:difftrans4}(c) shows a unique case which is most instructive. For $\varphi=\pi/2$ the coupling constant $\kappa=\kappa_1\pm\kappa_2, \in \mathbb{R}$ . Thus a maximum difference between $\eta_{-1}$ and $\eta_{+1}$ is obtained, culminating in the full depletion of $\eta_{-1}$ if $\kappa_1=\kappa_2$ (see appendix).  Finally, we want to draw the attention to the case of $\varphi=3\pi/4>\pi/2$. Then $\eta_{\pm1}$ gives identical results as for $\varphi-\pi/2$. The zero order diffraction efficiency $\eta_0$, however, approaches the mean absorption curve for $|\theta|\gg\theta_B$  from above in the case of  $\varphi<\pi/2$ and contrary from below for $\varphi>\pi/2$ .
Considering these arguments it is obvious, that only a simultaneous fit of all diffraction data, i.e., zero and $\pm$ first order diffracted intensities,  allows to extract the decisive parameters $\kappa_1,\kappa_2,\varphi$. On the other hand these curves are therefore fingerprints of the relation between the parameters. The following recipe can help in judging about the general situation (for $\alpha_0 d\approx 1$):
\begin{itemize}
\item Check $\eta_{\pm1}$: if their magnitudes differ, this is a  fingerprint that mixed gratings exist that are out of  phase ($\varphi\not=0$). The ratio $\eta_{+1}/\eta_{-1}$ at the Bragg position obtains a maximum value for $\varphi=\pi/2$ and for $\kappa_1=\kappa_2$ \cite{Ellabban-oex06}.
    \item Check $\eta_0$: if $\eta_0(\theta=0)<A(0)$ then $|\varphi|<\pi/2$ and else vice versa
    \item If $\eta_0^{max}+\eta_0^{min}>2A(\theta_B)$, the absorptive component is dominating and else vice versa.
    \item For overmodulated phase gratings another feature of the diffraction efficiencies becomes prominent: the side minima near the Bragg peak are lifted to nonzero values (for $\varphi\not=\pi/2$). This striking feature can already be understood in the case $\varphi=0$ where we simply add up the pure absorptive and the pure phase grating. The positions of the $s^{\rm th}$ side minima are then given by $\vartheta_s^{(1)}=2[(s\pi/D)^2-\kappa_1^2]^{1/2}$ (phase grating) and   $\vartheta_s^{(2)}=2[(s\pi/D)^2+\kappa_2^2]^{1/2}$ (absorption grating). Thus, their minima considerably deviate from each other for $\kappa_{1,2}\approx s\pi/D$. Recalling, that $\kappa_2=\alpha_1/2\leq\alpha_0/2$ such a situation will practically occur if $\kappa_1\gg\pi/D$, i.e., for $s>1$ (overmodulated phase gratings exist).  This is realized in various systems (see e.g., \cite{Neipp-jpd02,Neipp-joa01,Drevensek-Olenik-pre06} but did not deserve proper attention.
\end{itemize}
Finally, we would like to recall that for $\varphi\to\varphi+\pi$ the complex coupling parameters $\kappa^\pm\to\kappa^\mp$ are interchanged and thus the $\eta_{\pm1}\to\eta_{\mp 1}$. For $\eta_0$ the term in the second line of Equation \ref{eq:TE} changes sign because of $\cos{\varphi}\to\cos{(\varphi+\pi)}=-\cos{\varphi}$.
\section{Experimental and Discussion}
The investigations were performed on a pure congruently melted lithium niobate crystal (thickness: $5 {\rm mm}$). Holographic transmission gratings were prepared by a standard two-wave mixing setup using an argon-ion laser at a recording wavelength of $\lambda_p= 351$ nm. Two plane waves with equal intensities and parallel polarization states (s-polarization) were employed as recording beams under a crossing angle of $2\Theta_B = 20.21^\circ$ (outside the medium) corresponding to a grating period of 1000 nm where the polar  $c$-axis is lying in the plane of incidence. The total intensity of the writing beams was 9 mW/cm$^2$. 
HPDLC samples were fabricated from a UV curable mixture prepared from commercially available constituents as previously reported in literature \cite{Drevensek-Olenik-pre06}. The grating period was  1216 nm, the grating thickness  about $30 \mu$m \cite{Fally-prl06}.  After holographic recording we postcured the  sample by illuminating it homogeneously with one of the UV writing beams.

The grating characteristics of the samples was analyzed by monitoring the angular dependencies of the $\pm$ first and zero order diffraction efficiencies.  For this purpose the samples were fixed on an accurately controlled rotation stage with an accuracy of $\pm 0.001^\circ$) and facultatively (HPDLC) in a heating chamber. In the case of \linbo we used a single considerably reduced readout beam at $\lambda_r=\lambda_p=351$ nm and s-polarization, whereas for the HPDLC a He-Ne laser beam at a readout wavelength of $\lambda_r=543$ nm and p-polarization state was employed.
\begin{figure}
\includegraphics[width=\columnwidth]{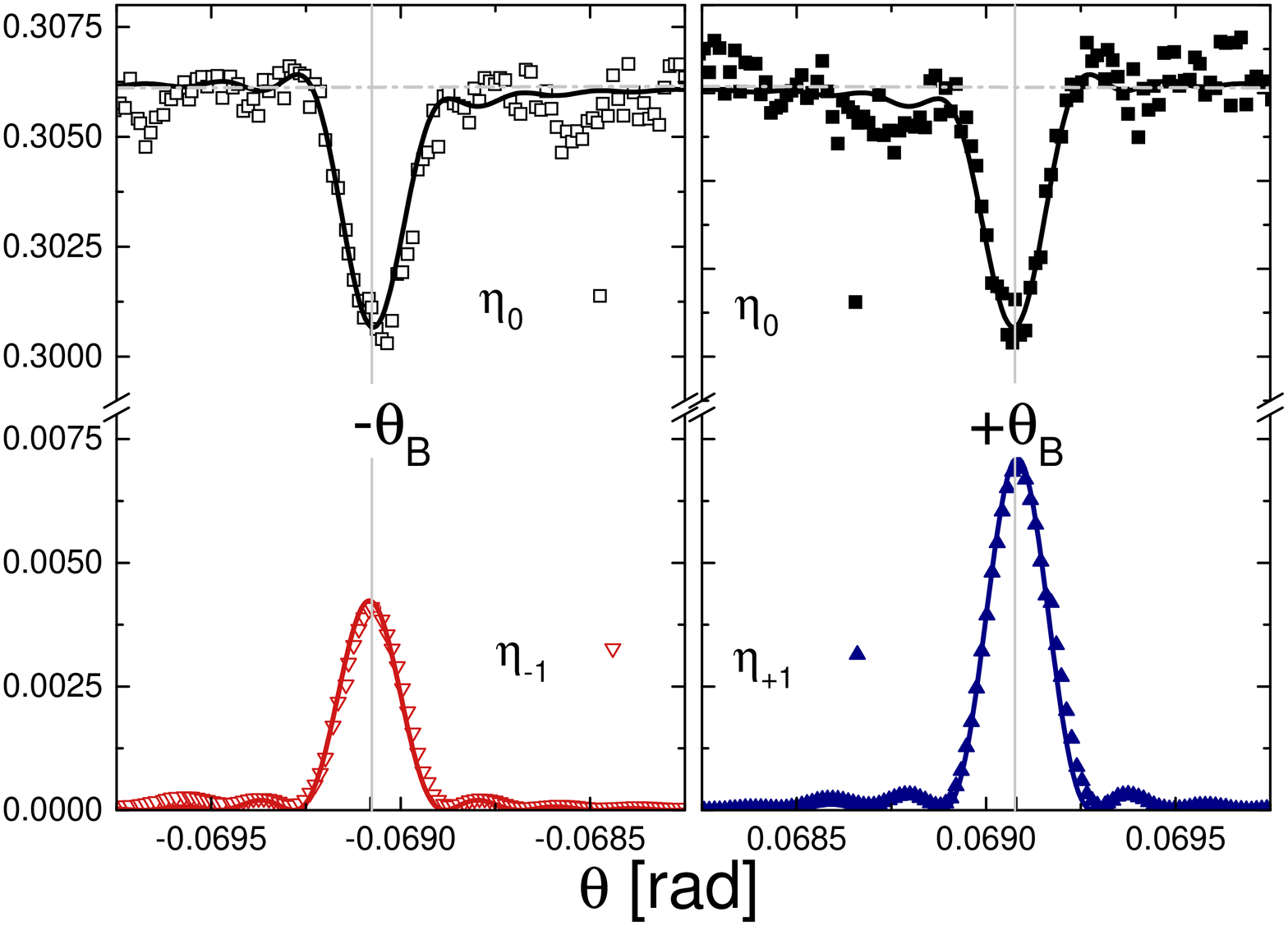}
\caption{\label{fig:linbo}
Angular dependence of the diffraction efficiencies for the zero and $\pm$ first orders of a grating recorded in a pure \linbo sample. The unequal values of the first order diffraction efficiencies are an impressive signature for the existence of mixed phase and amplitude gratings that are out of phase. The zero order diffraction efficiency at the Bragg angles show only a slight asymmetry with respect to $\theta_B$ because the diffraction efficiencies are small ($<1\%$) and the phase grating is by far dominating. The solid lines show a simultaneous fit to $\eta_{\pm1}$ and $\eta_0$. The dashed-dot line indicates the mean absorption curve.}
\end{figure}
Figure \ref{fig:linbo} shows the experimental curves for the $0., \pm1.$ diffraction orders from a grating recorded in nominally pure congruently melted \linbo. According to the recipe given above we immediately can diagnose  \emph{mixed out-of-phase} refractive-index and amplitude gratings, because the $\eta_{+1}>\eta_{-1}$. Further by inspecting the zero order diffraction we come to know that the phase $0<\varphi<\pi/2$. The effects in the zero order are not so prominent for two reasons: the overall diffraction efficiency is very small and the phase grating is dominant because the zero order diffraction curve extends mostly to values below the mean absorption curve (dash-dot line in Figure \ref{fig:linbo}). A simultaneous fit of Equations \ref{eq:DE} and \ref{eq:TE} to the measured data yielded the following parameters: $n_1=(3.01\pm0.04)\times 10^{-6}, \alpha_1=8.18\pm0.48 {\rm m}^{-1}, \varphi=1.027\pm0.059, \alpha_0=118\pm1.7{\rm m}^{-1}$ with a reduced chi-square value of $1.89\times10^{-7}$ . From this value and Figure \ref{fig:linbo} it is obvious that the equations excellently fit the data.

Finally, we intend to demonstrate the usability of the (qualitative) analysis employing an example with strong overmodulation and high extinction: holographic polymer-dispersed liquid crystals (H-PDLCs).
\begin{figure}[h]
\centering
\includegraphics[width=\columnwidth]{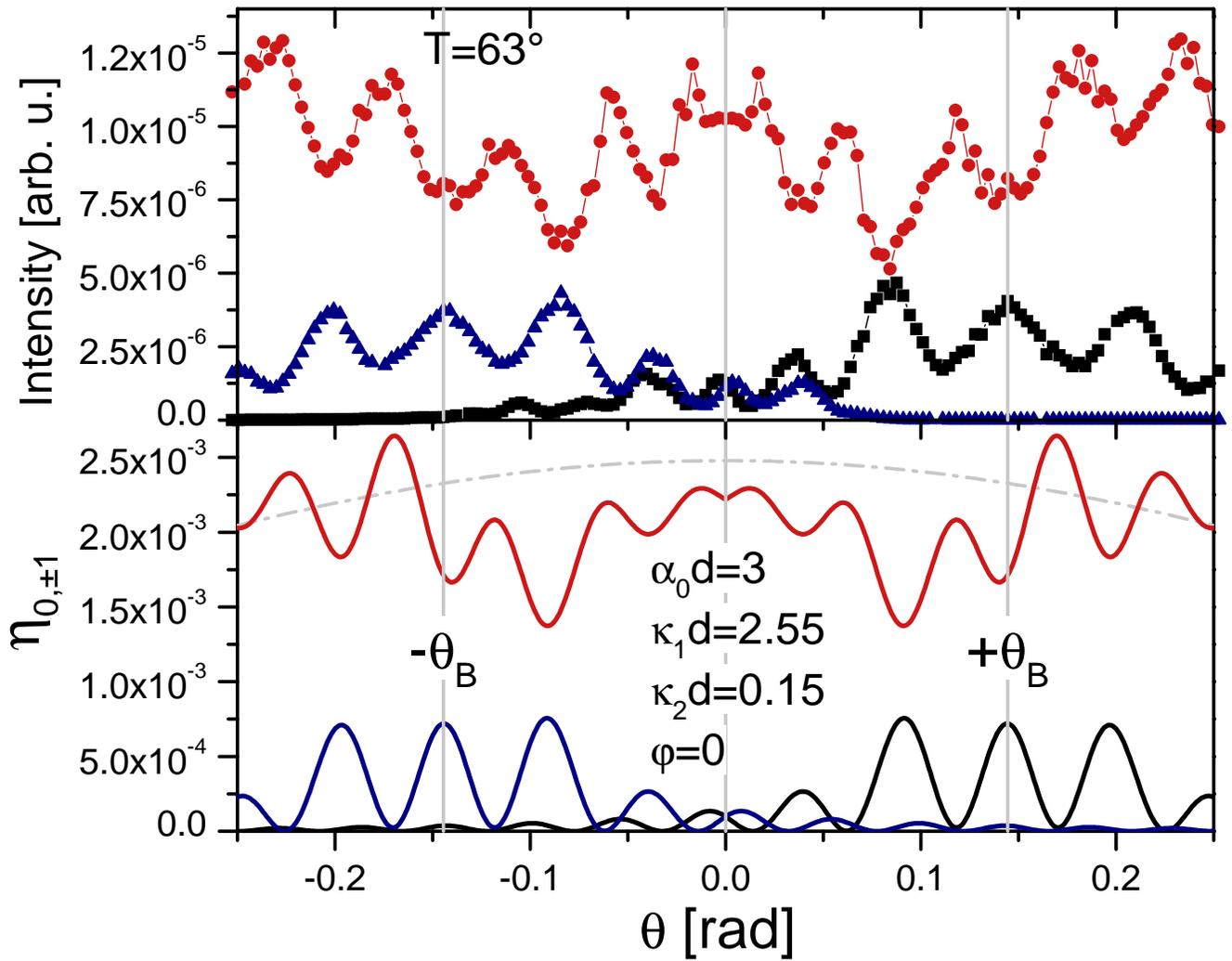}
\caption{\label{fig:hpdlc} Zeroth and $\pm$ first diffraction orders of a strongly overmodulated grating in HPDLC  at 63$^\circ$ Celsius. $\Lambda=1.2\mu$m, $\lambda$=543 nm, $d=30\quad \mu$m (same sample as used for the investigations in Ref. \cite{Drevensek-Olenik-pre06}. The lower graphs show a simulation according to Equations \ref{eq:DE} and \ref{eq:TE}. Note, that here the mean extinction is already rather high. We do not expect that a fit could be successful for at least three reasons: (1) as in HPDLCs  anisotropic gratings are formed, for the basic equations the full theory of Montemezzani and Zgonik should be employed \cite{Montemezzani-pre97}. (2) It can be noticed, that around $\theta=0$  more than two waves are propagating in the medium. Therefore, also the two-wave coupling theory is not fully applicable. Instead a rigorous coupled wave analysis should be performed \cite{Moharam-josa81}. (3) The gratings are expected to be inhomogeneous and non-sinusoidal \cite{Fally-prl06}, thus not completely fulfilling the requirements}
\end{figure}
Only recently was a preliminary beam-coupling analysis of such a system conducted, a task which is not simple from an experimental point of view \cite{Ellabban-spie07}, in particular if the experiments should be carried out under high temperatures or application of external electric fields.
Figure \ref{fig:hpdlc} shows the diffraction curves from a grating in a HPDLC  at an elevated temperature. We can understand  the major characteristic features as follows: The liquid crystal (LC) component in an HPDLC is highly birefringent. Statistical alignment of the LC-droplets of about the light wavelength's size leads to strong scattering, i.e., extinction which can be treated similar to absorption provided that multiple scattering does not play an essential role.  HPDLCs basically consist of alternating regions with high and low concentration of LCs embedded  in a polymer matrix. Thus, these periodically varying scatterers act as extinction gratings. In addition, of course, also the refractive index is strongly modulated (at least via the density changes). Therefore, HPDLCs are typical examples of mixed gratings.  Furthermore, it is well known in literature that the light-induced refractive-index changes are extremely high and strong overmodulation occurs (see e.g. \cite{Drevensek-Olenik-pre06}). Such an example is shown in Figure \ref{fig:hpdlc}. From  the experimental data we conclude, that combined refractive-index and extinction gratings are produced. This is consistent with our previous beam-coupling measurements \cite{Ellabban-spie07}.  However,  we do not dare to decide about a possible phase between them. A quantitative evaluation is not possible for this case as we are aware of the fact, that in HPDLCs the gratings are anisotropic and thus the basic equations of  Ref. \cite{Kogelnik-bell69} should be replaced by the full equations given by Montemezzani and Zgonik \cite{Montemezzani-pre97}.  In addition, the gratings are usually rather inhomogeneous across the sample but might be considerably improved upon further efforts during recording \cite{De-Sio-ao06}. The non-zero minima in the diffracted beams partially might originate from overmodulation as discussed above but mainly from the inhomogeneity of the gratings and a profile perpendicular to the grating vector \cite{Uchida-josa73}.
However, a qualitative understanding of the changes occuring during heating or applying an electric field can still be read off from the diffraction curves like those  shown in Ref. \cite{Drevensek-Olenik-pre06}.
    \section{Remarks and Conclusion}
The above discussed analysis is easily applicable for $\alpha_0d\approx 1,\kappa_1\approx\kappa_2$ and $\eta_1(\theta_B)/\eta_0(\theta_B)\gtrsim 0.01$, so that with the chosen example of \linbo above we are already at the limit. If one grating type is dominant the analysis still remains valid, however, the resulting values for $\varphi$ and the smaller component result in quite large errors.

We would like to draw the attention to the fact, that for $\alpha_0\ll1$ the absorptive grating strength is considerably limited, so that in general the zero order diffraction will not feel the Bragg diffraction. On the other hand, for $\alpha_0d\gg1$,  the forward diffracted beam will exhibit a maximum near the Bragg position, a fact which is well known in x-ray optics (anomalous transmission), see e.g. \cite{Batterman-rmp64}.

We would like to point out, that the analysis of only the first diffraction orders cannot give the full information on all relevant parameters \cite{Ellabban-oex06}. However, it is sufficient to use the $\pm$ first together with the zero order diffraction and to avoid more demanding beam-coupling (interferometric) experiments. A prospective phase between the grating and the interference pattern\cite{Sutter-josab90,Kahmann-josaa93,Fally-josaa06}, however, cannot be determined by simple diffraction experiments.

We further would like to emphasize, that  the limitations of the coupled wave equations according to Ref. \cite{Kogelnik-bell69} should be kept in mind when employing  Equations \ref{eq:DE} and \ref{eq:TE}, e.g., it is assumed that the gratings are planar, purely sinusoidal and isotropic (for anisotropic gratings the theory given in Ref. \cite{Montemezzani-pre97} should be employed), $\alpha_1\leq\alpha_0$ (for violation of this condition see \cite{Shcheulin-os07}) and only two beams are kept in the coupling scheme. If the latter is not applicable the theory of rigorous coupled waves has to be applied \cite{Moharam-josa81}, naturally with an increase of  the number of coupling constants between the beams and thus with loss of simplicity.

\section*{Acknowledgment}
We are grateful to Profs. Th. Woike and M. Imlau for providing the \linbo sample. Financially supported by the Austrian Science Fund (P-18988) and the \"OAD in the frame of the STC program Slovenia-Austria (SI-A4/0708). We acknowledge continuous support by E. Tillmanns by making one of his labs available to us.
\section*{Appendix}
For the particular case of $\varphi=\pi/2$ the diffraction efficiencies read:
\begin{eqnarray}
\eta_{\pm1}(\theta)&=&A(\theta)z^{-1} 4 (\kappa_1\pm\kappa_2)^2\sin^2{\left(z^{1/2}D/2\right)}= \nonumber\\
&=&A(\theta)z^{-1} r^{\pm1} 4 (\kappa_1^2-\kappa_2^2)\sin^2{\left(z^{1/2}D/2\right)}\\
\eta_0&=&A(\theta)z^{-1}\left[\vartheta^2+4(\kappa_1^2-\kappa_2^2) \cos^2{\left(z^{1/2}D/2\right)}\right]\\
z&=&\vartheta^2+4(\kappa_1^2-\kappa_2^2)\nonumber.
\end{eqnarray}
It's interesting to note, that for this case the diffracted and forward diffracted beams have the functional dependence of  pure phase gratings with an effective coupling constant of $2[\kappa_1^2-\kappa_2^2]^{1/2}$. The amplitude of the diffracted beams, however, is enhanced or diminished by a multiplication with or division by  $r=(\kappa_1-\kappa_2)/(\kappa_1+\kappa_2)$, respectively. Therefore, it's easy to see that for $\kappa_1=\kappa_2$ the curves shown in Figure \ref{fig:difftrans4} (c) arise.

\end{document}